\documentclass[conference]{IEEEtran}
\usepackage[dvips]{graphicx}
\usepackage[cmex10]{amsmath}
\usepackage{multirow}
\usepackage{epsfig}
\usepackage{mathrsfs}
\usepackage{amssymb}
\usepackage{comment}
\usepackage{bm}
\usepackage{url}
\usepackage{array}
\usepackage{amsthm}
\usepackage{blkarray}
\usepackage{fancyhdr}
\usepackage{enumerate}
\usepackage[lined,boxed,commentsnumbered,ruled,linesnumbered]{algorithm2e}

\renewcommand{\thispagestyle}[1]{} % do nothing

\theoremstyle{definition}
\newtheorem{theorem}{Theorem}[section]
\newtheorem{lemma}[theorem]{Lemma}

\ifCLASSINFOpdf

\else

\fi

\usepackage[tight,footnotesize]{subfigure}

\begin{document}
\pagestyle{fancy}
\IEEEoverridecommandlockouts

\lhead{\textit{Technical Report, Dept. of EEE, Imperial College, London, UK, Feb., 2013.}}
\rhead{} %\thepage can be added to the brackets
\title{Link Identifiability with Two Monitors: Proof of Selected Theorems}
\author{\IEEEauthorblockN{Liang Ma\IEEEauthorrefmark{2}, Ting He\IEEEauthorrefmark{3}, Kin K. Leung\IEEEauthorrefmark{2}, Ananthram Swami\IEEEauthorrefmark{4}, and Don Towsley\IEEEauthorrefmark{1}}
\IEEEauthorblockA{\IEEEauthorrefmark{2}Imperial College, London, UK. Email: \{l.ma10, kin.leung\}@imperial.ac.uk\\
\IEEEauthorrefmark{3}IBM T. J. Watson Research Center, Hawthorne, NY, USA. Email: the@us.ibm.com\\
\IEEEauthorrefmark{4}Army Research Laboratory, Adelphi, MD, USA. Email: ananthram.swami.civ@mail.mil\\
\IEEEauthorrefmark{1}University of Massachusetts, Amherst, MA, USA. Email: towsley@cs.umass.edu
}
%\thanks{Research was sponsored by the U.S. Army Research Laboratory and the
%U.K. Ministry of Defence and was accomplished under Agreement Number
%W911NF-06-3-0001. The views and conclusions contained in this document
%are those of the authors and should not be interpreted as representing the
%official policies, either expressed or implied, of the U.S. Army Research
%Laboratory, the U.S. Government, the U.K. Ministry of Defence or the U.K.
%Government. The U.S. and U.K. Governments are authorized to reproduce and
%distribute reprints for Government purposes notwithstanding any copyright
%notation hereon.}
}

\maketitle

\IEEEpeerreviewmaketitle

\section{Introduction}
Selected lemmas and theorems in \cite{MaGlobecom} are proved in detail in this report. We first list the theorems in Section II and then give the corresponding proofs in Section III. See the original paper \cite{MaGlobecom} for terms and definitions. Table~\ref{t notion} summarizes all graph-theoretical notions used in this report (following the convention in \cite{GraphTheory2005}).

\begin{table}[tb]
\vspace{.5em}
\renewcommand{\arraystretch}{1.3}
\caption{Notations in Graph Theory} \label{t notion}
\vspace{-.5em}
\centering
\begin{tabular}{r|m{6.26cm}}
  \hline
  \textbf{Symbol} & \textbf{Meaning} \\
  \hline
  $V(\mathcal{G})$, $L(\mathcal{G})$ & set of nodes/links in graph $\mathcal{G}$\\
  \hline
  $|\mathcal{G}|$ & degree of $\mathcal{G}$: $|\mathcal{G}|=|V(\mathcal{G})|$ (number of nodes)\\
  \hline
  $||\mathcal{G}||$ & order of $\mathcal{G}$: $||\mathcal{G}||=|L(\mathcal{G})|$ (number of links)\\
  \hline
  $\mathcal{G} \cup \mathcal{G}^{'} $ & union of graphs: $ \mathcal{G} \cup \mathcal{G}^{'}=(V \cup V^{'}, L \cup L^{'})$\\
  \hline
  $\mathcal{H}$ & interior graph\\
  \hline
  $\mathcal{P}$ & simple path\\
  \hline
  $m_i$ & $m_i\in V(\mathcal{G})$ is the $i$-th ($i=\{1,2\}$) monitor in $\mathcal{G}$ \\
  \hline
  $W_l$, $W_{\mathcal{P}}$ &  metric on link $l$ and sum metric on path $\mathcal{P}$ \\
  \hline
  $m'_1$, $m'_2$ & two monitoring agents in a biconnected component\\
  \hline
\end{tabular}
\vspace{-.5em}
\end{table}

\section{Theorems}

\begin{lemma}
\label{lemma:effectiveMonitor}
Let $\mathcal{B}$ be a biconnected component with monitoring agents $m'_1$ and $m'_2$. The set of identifiable links in $L(\mathcal{B})$ does not depend on whether $m'_1$ or $m'_2$ are monitors or not, except for link $m'_1m'_2$ (if it exists).  Link $m'_1m'_2$ is identifiable if and only if $m'_1$ and $m'_2$ are both monitors.
\end{lemma}

\begin{theorem}
\label{theorem:FLIcorrectness}
Algorithm DIL-2M, Determining Identifiable Links under Two Monitors, can determine all identifiable links in a network with given 2-monitor placement.
\end{theorem}

\section{Proofs}

\subsection{Proof of Lemma~\ref{lemma:effectiveMonitor}}
\begin{figure}[tb]
\centering
\includegraphics[width=2.1in]{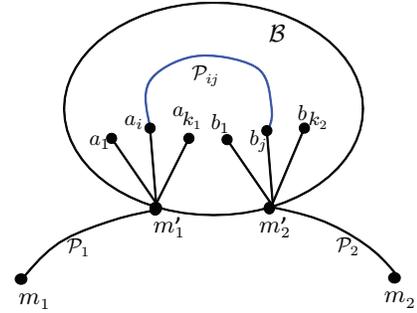}
\vspace{-.5em}
\caption{Monitoring agents $m'_1$ and $m'_2$ wrt biconnected component $\mathcal{B}$.}\label{fig:EffectiveMonitorProof}
\vspace{-.5em}
\end{figure}

\emph{1)} Let $m'_1$ and $m'_2$ be the two monitoring agents of biconnected component $\mathcal{B}$ in Fig.~\ref{fig:EffectiveMonitorProof} and $m'_1$ ($m'_2$) connects to the real monitor $m_1$ ($m_2$) by path\footnote{$\mathcal{P}_1$ and $\mathcal{P}_2$ may not be unique.} $\mathcal{P}_1$ ($\mathcal{P}_2$), i.e., none of $m'_1$ and $m'_2$ are real monitors. In Fig.~\ref{fig:EffectiveMonitorProof}, it is impossible that $\mathcal{P}_1$ and $\mathcal{P}_2$ must have a common node; since otherwise $m'_1$ and $m'_2$ are not cut-vertices, contradicting the processing of localizing monitoring agents for a biconnected component (see DIL-2M). To identify link metrics in $\mathcal{B}$, all measurement paths involving links in $\mathcal{B}$ are of the following form
\begin{equation}\label{eq:effectiveMonitorEqForm}
    W_{\mathcal{P}_1}+W_{m'_1a_i}+W_{\mathcal{P}_{ij}}+W_{b_jm'_2}+W_{\mathcal{P}_2}=c'_{ij},
\end{equation}
assuming $\mathcal{P}_1$ ($\mathcal{P}_2$) is always selected to connect $m_1$ and $m'_1$ ($m_2$ and $m'_2$). We know that if $m'_1$ and $m'_2$ are real monitors, then each measurement (except direct link $m'_1m'_2$) path is of form
\begin{equation}\label{eq:MonitorEqForm}
    W_{m'_1a_i}+W_{\mathcal{P}_{ij}}+W_{b_jm'_2}=c_{ij}.
\end{equation}
Therefore, compared with (\ref{eq:MonitorEqForm}), (\ref{eq:effectiveMonitorEqForm}) is equivalent to abstracting each of ${\mathcal{P}_1}+{m'_1a_i}$ and ${b_jm'_2}+{\mathcal{P}_2}$ as a single link. By Theorem~III.1 \cite{MaGlobecom}, we know that none of the  exterior links are identifiable. Thus, the link metrics of exterior links do not affect the identification of interior links. Therefore, $\mathcal{B}$ can be visualized as a network with two monitors $m'_1$ and $m'_2$ but each exterior link in $\{\{m'_1a_i\},\{m'_2b_j\}\}$ has an added weight from $W_{\mathcal{P}_1}$ or $W_{\mathcal{P}_2}$. The above argument also holds when $m_1$ ($m_2$) chooses another path, say $\mathcal{P}'_1$ ($\mathcal{P}'_2$), to connect to $m'_1$ ($m'_2$), then it simply implies that different exterior links in $\{\{m'_1a_i\},\{m'_2b_j\}\}$ in $\mathcal{B}$ may have different added path weights when regarding $m'_1$ and $m'_2$ as two monitors. Moreover, the above conclusion also applies to the case that one of $m'_1$ and $m'_2$ is a real monitor. Therefore, the identifiability of all links except for the direct link $l_d=m'_1m'_2$ (if any) remains the same regardless whether $m'_1$, $m'_2$ are monitors or not.\looseness=-1

\emph{2)} To identify direct link (if any) $l_d:=m'_1m'_2$, all measurement paths traversing $l_d$ must utilize unidentifiable links incident to $m_1$ or $m_2$. To eliminate these unidentifiable links in linear equations, some other measurement paths in $\mathcal{B}$ must be used; however, each measurement path in $\mathcal{B}$ introduces two new uncomputable variables $W_{m'_1a_i}$ and $W_{b_jm'_2}$, and thus each newly added path for identifying $l_d$ involves new unknown variables. Therefore, $l_d$ cannot be identified when one of $m'_1$ and $m'_2$ is not a real monitor, i.e., $m'_1$ and $m'_2$ must be both real monitors such that $l_d$ is identifiable.
\hfill$\blacksquare$

\subsection{Proof of Theorem~\ref{theorem:FLIcorrectness}}
\emph{1) completeness of four categories.}
DIL-2M only processes the biconnected components with 2 monitoring agents as none of the links in biconnected components with 1 or 0 monitoring agent are identifiable. Since only 2 monitors are used in $\mathcal{G}$, the number of monitoring agents for each biconnected component cannot be greater than 2; therefore, it is correct for DIL-2M to only process the biconnected components with 2 monitoring agents. If a triconnected component contains only a single link, then this triconnected component is also a biconnected component, whose identifiability is determined by line 2-4 in DIL-2M based on Lemma~\ref{lemma:effectiveMonitor}. Therefore, the four identification categories do not consider the case of a triconnected component which is a single link. Now we discuss triconnected components (with at least 3 nodes) as follows.

\begin{figure}[tb]
\centering
\includegraphics[width=1.2in]{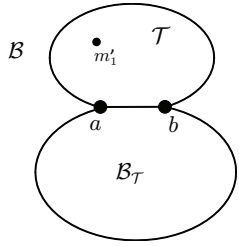}
\vspace{-.5em}
\caption{Triconnected component $\mathcal{T}$ in biconnected component $\mathcal{B}$, where $\{a,b\}$ is the 2-vertex-cut, $\mathcal{B}_\mathcal{T}$ is the neighboring biconnected component connecting to $\mathcal{T}$ via $\{a,b\}$ and $m'_1$ is a monitoring agent.}\label{fig:TriOnePP}
\vspace{-.5em}
\end{figure}

\emph{(i) A triconnected component $\mathcal{T}$ containing only one monitoring agent.} In this case, $\mathcal{T}$ must contain one 2-vertex-cut as $\mathcal{T}$ contains 2 monitoring agents otherwise. This case is illustrated in Fig.~\ref{fig:TriOnePP}, where $\{a,b\}$ is the 2-vertex-cut, $\mathcal{T}_N$ is the neighboring biconnected component connecting to $\mathcal{T}$ via $\{a,b\}$ and $m'_1$ is a monitoring agent. Since the associated biconnected component contains two monitoring agents, the neighboring component $\mathcal{T}_N$ must contain one monitoring agent, which cannot be the same as $a$ or $b$ as $\mathcal{T}$ involves two monitoring agents otherwise. Thus, $\{a,b\}$ is of Type-1-VC. If $m'_1\notin \{a,b\}$, then $\mathcal{T}$ belongs to Category 1. If $m'_1=a$ or $m'_1=b$, then $\mathcal{T}$ belongs to Category 2.

\emph{(ii) A triconnected component $\mathcal{T}$ containing two monitoring agents.} Obviously, this triconnected component is of Category 2.
\begin{figure}[tb]
\centering
\includegraphics[width=3.1in]{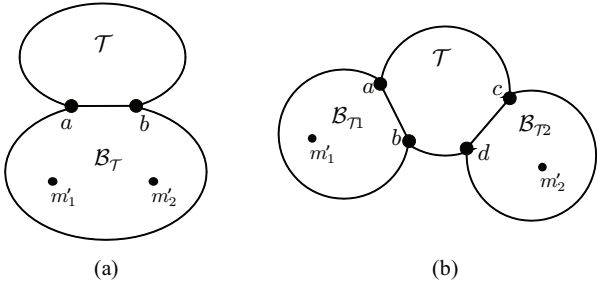}
\vspace{-.5em}
\caption{Triconnected component containing no monitoring agents.}\label{fig:TriNoPP}
\vspace{-.5em}
\end{figure}

\emph{(iii) A triconnected component $\mathcal{T}$ containing no monitoring agents.} This case can be further divided into two sub-cases, i.e., Fig.~\ref{fig:TriNoPP}-a (monitoring agents reside in the same neighboring biconnected component) and Fig.~\ref{fig:TriNoPP}-b (monitoring agents reside in two neighboring biconnected components). For Fig.~\ref{fig:TriNoPP}-a, there exist two monitoring agents $m'_1$ and $m'_2$ in the neighboring component $\mathcal{B}_\mathcal{T}$ within the same parent biconnected component, thus resulting $\mathcal{T}$ to be of Category 2. Fig.~\ref{fig:TriNoPP}-b illustrates the case that each neighboring component ($\mathcal{B}_{\mathcal{T}1}$ and $\mathcal{B}_{\mathcal{T}2}$) contains one monitoring agent. According to the connectivity of $\mathcal{T}$ in Fig.~\ref{fig:TriNoPP}-b, $\mathcal{T}$ belongs to either Category 3 or 4.

Therefore, excluding the triconnected component containing a single link, Category 1-4 are complete to cover all cases of triconnected component within biconnected components with 2 monitoring agents.

\emph{2) identification of each category.}
In Theorem III.2 \cite{MaGlobecom}, the prerequisite for network identifiability is that all involved links can be used for constructing measurement paths. In DIL-2M, we sequentially consider each triconnected component which possibly contains virtual links (see \cite{MaGlobecom}). These virtual links, however, do not exist in real networks. To tackle with this issue, we have the following Claim.

\textbf{Claim 1.} A triconnected component $\mathcal{T}$ may contain multiple virtual links. For each involved virtual link whose end-points $\{v_1,v_2\}$ (the end-points of a virtual link must form a vertex cut) are neither Type-1-VC nor Type-2-VC (used to determine\footnote{If $\mathcal{T}$ contains one Type-1-VC $\{a,b\}$ and $a$ (or $b$) is a monitoring agent, then this Type-1-VC $\{a,b\}$ is not used to determine the category of $\mathcal{T}$. This is because, within $\mathcal{T}$, there must exist another monitoring agent $m'$ ($m'\neq a\neq b$) or aother Type-1-VC which is used to determine the category of $\mathcal{T}$ since the parent biconnected component of $\mathcal{T}$ contains two monitoring agents.} the category of $\mathcal{T}$) wrt $\mathcal{T}$, there exists a simple path $\mathcal{P}_r$ with the same end-points in a neighboring biconnected component which connects to $\mathcal{T}$ via $\{v_1,v_2\}$. $\mathcal{P}_r$ can be used to replace the associated virtual link in $\mathcal{T}$ if this virtual link is chosen to construct measurement paths for identifying real links in $\mathcal{T}$. This replacement operation does not affect all existing path construction policies or the identification properties of real links in $\mathcal{T}$.
\begin{figure}[tb]
\centering
\includegraphics[width=2.4in]{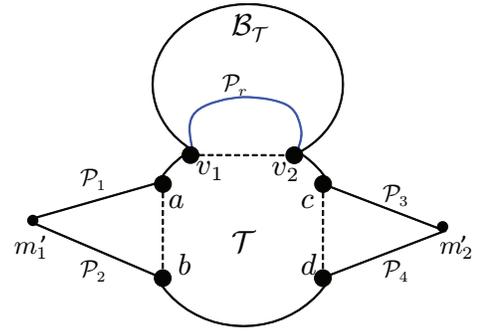}
\vspace{-.5em}
\caption{Virtual link replacement.}\label{fig:Claim1}
\vspace{-.5em}
\end{figure}

\begin{proof}
Fig.~\ref{fig:Claim1} illustrates a triconnected component $\mathcal{T}$ with two Type-1-VCs $\{a,b\}$ and $\{c,d\}$. For vertex cut $\{v_1,v_2\}$ (which is neither $\{a,b\}$ nor $\{c,d\}$), there exists a simple path $\mathcal{P}_r$ connecting $v_1$ and $v_2$ in the neighboring biconnected component $\mathcal{B}_\mathcal{T}$ of $\mathcal{T}$ as $\mathcal{B}_\mathcal{T}$ contains at least 3 nodes. We know that $\mathcal{B}_\mathcal{T}$ connects to the $\mathcal{T}$-involved component by only $\{v_1,v_2\}$; therefore, $\mathcal{P}_1$, $\mathcal{P}_2$, $\mathcal{P}_3$ and $\mathcal{P}_4$ do not have common nodes with $\mathcal{B}_\mathcal{T}$ except $a$ ($c$) might equal $v_1$ ($v_2$). Hence, for virtual link $v_1v_2$, if it is used for identifying real links in $\mathcal{T}$ based on Theorem III.2 \cite{MaGlobecom}, then it can be replaced by $\mathcal{P}_r$ which is a simple path and can be abstracted as a real link in $\mathcal{T}$. In the cases that $\mathcal{T}$ contains other combinations of Type-1-VCs and Type-2-VCs, the same argument applies.
\end{proof}
Now we discuss the link identifications of the four categories in DIL-2M.

\begin{figure}[tb]
\centering
\includegraphics[width=1.9in]{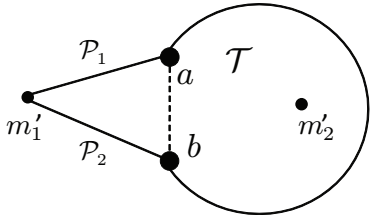}
\vspace{-.5em}
\caption{Link identifications of Category 1.}\label{fig:Category1}
\vspace{-.5em}
\end{figure}
(i) Category 1.

Fig.~\ref{fig:Category1} illustrates the case of Category 1, where $m'_1$ and $m'_2$ are monitoring agents. Since the identification of $\mathcal{T}$ is within a biconnected component, there exist two paths $\mathcal{P}_1$ and $\mathcal{P}_2$. $\mathcal{P}_1$ and $\mathcal{P}_2$ are internally vertex disjoint, since if $\mathcal{P}_1$ and $\mathcal{P}_2$ must have a common node (except $m'_1$), then the common node is a cut vertex, contradicting the property of biconnectivity. With Claim 1, all virtual links in $\mathcal{T}$ (except $ab$) can be replaced by the corresponding real paths in neighboring components; therefore, we only need to consider one virtual link $ab$ (if any) in Fig.~\ref{fig:Category1}. Abstracting $m'_1$ and $m'_2$ as two monitors (by Lemma~\ref{lemma:effectiveMonitor}) and $\mathcal{P}_1$ and $\mathcal{P}_2$ as two links, Fig.~\ref{fig:Category1} satisfies the interior graph identifiability conditions (Theorem III.2 \cite{MaGlobecom}) and exterior link unidentifiability conditions (Theorem III.1 \cite{MaGlobecom}) naturally if there exists real link $ab$. Now we consider that there is no link $ab$ in the original graph, i.e., $ab$ is a virtual link, and still abstract $\mathcal{P}_1$ and $\mathcal{P}_2$ as two links for identifying $\mathcal{T}$. If $\mathcal{T}$ is a triangle, then the two exterior links ($am'_2$ and $bm'_2$ if any) are unidentifiable according to Theorem III.1 \cite{MaGlobecom}. Now consider the case that $\mathcal{T}$ is 3-vertex-connected. Deleting any two links\footnote{The link can be a path from neighboring biconnected component for virtual link replacement.} in $\mathcal{T}$, the resulting graph is connected as the 3-vertex-connectivity of $\mathcal{T}$ implies the 3-edge-connectivity. Deleting $\mathcal{P}_1$ (or $\mathcal{P}_2$) and one link in $\mathcal{T}$, we also get a connected remaining graph. Thus, Fig.~\ref{fig:Category1} satisfies Condition \textcircled{\small 1} \normalsize in Theorem III.2 \cite{MaGlobecom}. Now consider deleting some vertices in $\mathcal{T}$. Deleting any two nodes in $\mathcal{T}$, the remaining graph of $\mathcal{T}$ is still connected as $\mathcal{T}$ is 3-vertex-connected. In this case, if $m'_1$ is isolated ($a$ and $b$ are deleted), $m'_1$ can reconnect to the remaining part of $\mathcal{T}$ by added link $m'_1m'_2$. When deleting $m'_1$ and any node in $\mathcal{T}$, the remaining graph of Fig.~\ref{fig:Category1} is obviously connected, thus satisfying Condition \textcircled{\small 2} \normalsize in Theorem III.2 \cite{MaGlobecom}. Therefore, for $\mathcal{T}$, all real links incident to $m'_2$ are unidentifiable and the rest links are identifiable.

\begin{figure}[tb]
\centering
\includegraphics[width=2.5in]{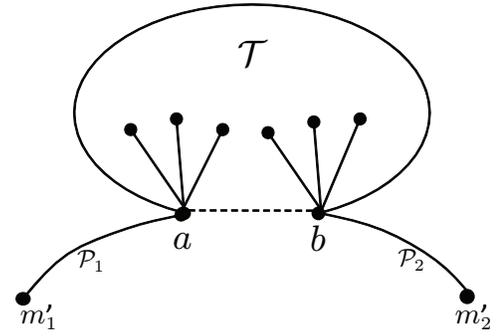}
\vspace{-.5em}
\caption{Link identifications of Category 2.}\label{fig:Category2}
\vspace{-.5em}
\end{figure}
(ii) Category 2.

Suppose $\mathcal{T}$ contains one Type-2-VC, no monitoring agents and no real link $ab$, as illustrated Fig.~\ref{fig:Category2}. Then there exist vertex disjoint paths $\mathcal{P}_1$ and $\mathcal{P}_2$ in the corresponding biconnected components. Thus, it is equivalent to the case in Fig.\ref{fig:EffectiveMonitorProof}. We also have that $\mathcal{T}$ is a triconnected component. Therefore, for $\mathcal{T}$, all real links incident to $a$ and $b$ are unidentifiable and the rest links are identifiable. On top of the discussed scenario shown in Fig.~\ref{fig:Category2}, now we consider the case that there exists real link $ab$ in $\mathcal{T}$, i.e., $ab$ is a real link in Fig.~\ref{fig:Category2}. Since $\{a,b\}$ is a Type-2-VC wrt $\mathcal{T}$ and none of $a$ or $b$ are monitoring agents, $\{a,b\}$ is \emph{not} a Type-1-VC or Type-2-VC or two monitoring agents wrt the neighboring components connecting to $\mathcal{T}$ via $\{a,b\}$. Thus, in those neighboring components, $ab$ is only an ordinary link (not incident to the two vertices of a Type-1-VC or Type-2-VC), the identifiability of which can be determined in identifying the neighboring components. Therefore, we do not need to consider the identifiability of $ab$ in the current triconnected component $\mathcal{T}$.

Suppose $\mathcal{T}$ contains one Type-2-VC and one monitoring agent, i.e., $a=m'_1$ or $b=m'_2$. Then it is still equivalent to the case in Fig.\ref{fig:EffectiveMonitorProof} for identifying links (except direct link $ab$) in $\mathcal{T}$. Since one of $a$ or $b$ is a monitoring agent, the link $ab$ incident to this monitoring agent is unidentifiable according to Theorem III.1 \cite{MaGlobecom} unless both end-points of $ab$ are monitoring agents (Lemma~\ref{lemma:effectiveMonitor}), which contradicts the assumption that $\mathcal{T}$ contains only one monitoring agent.

Suppose $\mathcal{T}$ contains two monitoring agents. Then $\mathcal{T}$ itself satisfies the interior graph identifiability conditions (Theorem III.2 \cite{MaGlobecom}) and exterior link unidentifiability conditions (Theorem III.1 \cite{MaGlobecom}). For the direct link connecting to these two monitoring agents, the identifiability is already determined by line 2-4 of DIL-2M.

In sum, for Category 2, the effective interior (exterior) links are identifiable (unidentifiable) and the effective direct link is determined in the identification of other triconnected components.

(iii) Category 3.

Category 3 is illustrated in Fig.~\ref{fig:Claim1}. Due to the 2-vertex-connectivity of the corresponding biconnected component, there exist pairwise internally vertex disjoint paths $\mathcal{P}_1$, $\mathcal{P}_2$, $\mathcal{P}_3$ and $\mathcal{P}_4$. Then similar arguments for discussing Category 1 can be applied. Therefore, $\mathcal{T}$ is the effective interior graph of $\mathcal{T}\cup\mathcal{P}_1\cup\mathcal{P}_2\cup\mathcal{P}_3\cup\mathcal{P}_4$ and thus all involved links in $\mathcal{T}$ are identifiable when $\mathcal{T}$ is 3-vertex-connected.

\begin{figure}[tb]
\centering
\includegraphics[width=2.4in]{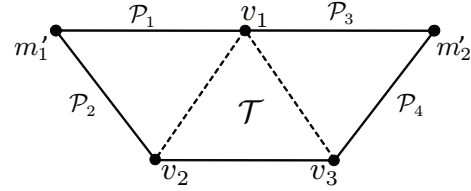}
\vspace{-.5em}
\caption{Link identifications of Category 4.}\label{fig:Category4}
\vspace{-.5em}
\end{figure}

(iv) Category 4 (processed by auxiliary algorithm - Algorithm 3).

Let the two Type-1-VCs be $\{v_1,v_2\}$ and $\{v_1,v_3\}$ and $S_1$ ($S_2$) the set of immediately neighboring triconnected components connecting to $\mathcal{T}$ via $\{v_1,v_2\}$ ($\{v_1,v_3\}$), as illustrated in Fig.~\ref{fig:Category4}. Based on the above discussions for Category 1-3, we know that there exist internally vertex disjoint paths $\mathcal{P}_1$, $\mathcal{P}_2$, $\mathcal{P}_3$ and $\mathcal{P}_4$. If $v_1v_2$ (or $v_1v_3$) is a real link, then $v_1v_2$ (or $v_1v_3$) is known as a Cross-link \cite{Ma13IMC} since $v_2v_3$ can be replaced by a path in neighboring biconnected component if $v_2v_3$ is virtual. Therefore, $v_1v_2$ (or $v_1v_3$) is identifiable. Now we focus on the identification of $v_2v_3$ in $\mathcal{T}$ (when $v_2v_3$ is a real link). The conditions to guarantee the identifiability of $v_2v_3$ is:
\\(link $v_1v_2$ is real \textbf{OR} $|S_1|\geq 2$ \textbf{OR} one component in $S_1$ is 3-vertex-connected) \textbf{AND} (link $v_1v_3$ is real \textbf{OR} $|S_2|\geq 2$ \textbf{OR} one component in $S_2$ is 3-vertex-connected).
\\Suppose the above condition is not satisfied. Then we can prove $v_2v_3$ is unidentifiable as follows.

\begin{figure}[tb]
\centering
\includegraphics[width=3.5in]{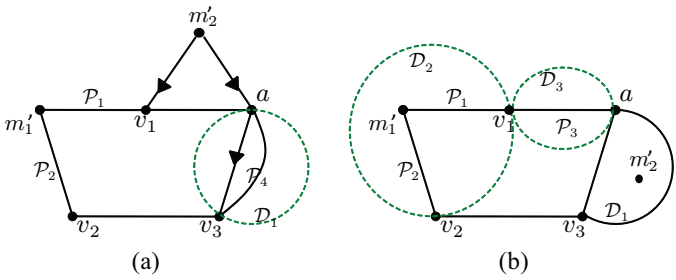}
\vspace{-.5em}
\caption{Category 4 - unidentifiable $v_2v_3$.}\label{fig:Category4Unidentifiable}
\vspace{-.5em}
\end{figure}

We first consider the condition: link $v_1v_3$ is real \textbf{OR} $|S_2|\geq 2$ \textbf{OR} one component in $S_2$ is 3-vertex-connected. If not satisfied, then it means there is no real link $v_1v_3$ and the only immediately neighboring triconnected component is a triangle, i.e., $v_1$-$a$-$v_3$ in Fig.~\ref{fig:Category4Unidentifiable} ($v_1a$ and $v_3a$ can be virtual links as well).

(iv-a) If the monitoring agent $m'_2$ is in the location shown in Fig.~\ref{fig:Category4Unidentifiable}-a, then all paths from $m'_1$ to $m'_2$ traversing $v_2v_3$ must use one simple path in $\mathcal{D}_1$. Therefore, the best case is that we can compute the sum metric of link ${v_2v_3}$ and another link which is incident to $v_3$ in $\mathcal{D}_1$, but cannot compute them separately.

(iv-b) If the monitoring agent $m'_2$ is in the location shown in Fig.~\ref{fig:Category4Unidentifiable}-b, then $\mathcal{P}_3$ and $v_2v_3$ become a ``double bridge'' connecting $\mathcal{D}_2$ and $\mathcal{D}_1$. Abstracting $\mathcal{P}_3$ as a single link, \cite{Ma13IMC} proves that none of the links in a double bridge is identifiable when constraining the measurement paths to simple paths. If we choose other paths as $\mathcal{P}_3$ in $\mathcal{D}_3$, then the same argument applies. Therefore, based on (iv-a) and (iv-b), $v_2v_3$ is unidentifiable.

Analogously, we can prove that $v_2v_3$ is unidentifiable when condition (link $v_1v_2$ is real \textbf{OR} $|S_1|\geq 2$ \textbf{OR} one component in $S_1$ is 3-vertex-connected) is not satisfied.

When the required conditions are satisfied, we can prove that $v_2v_3$ is identifiable as follows:

\vspace{-.5em}
If $v_1v_2$ ($v_1v_3$) is a virtual link, then it can be replaced by a path in a neighboring component. For instance, if $|S_1|\geq 2$, then one replacement path can be found in one component of $S_1$. If one component in $S_1$ is 3-vertex-connected, then there exist 2 internally vertex disjoint paths (each with the order greater than 1) connecting $v_1$ and $v_2$. Thus, we can choose one of them as a replacement path. Note that the virtual links possibly involved in the replacement paths can be further replaced by the paths in their neighboring components recursively. After these replacement operations, $v_2v_3$ in Fig.~\ref{fig:Category4} is a Shortcut (defined in \cite{Ma13IMC}), which is proved to be identifiable in \cite{Ma13IMC}.

Therefore, the auxiliary algorithm (Algorithm 3) of DIL-2M can determine all identifiable/unidentifiable links in a triangle triconnected component.

Consequently, with the complete coverage of four categories and the identification efficacy of each category, DIL-2M can determine all identifiable/unidentifiable links.
\hfill$\blacksquare$

\bibliographystyle{IEEEtran}
\bibliography{mybibSimplified}

\begin{thebibliography}{1}
\providecommand{\url}[1]{#1}
\csname url@samestyle\endcsname
\providecommand{\newblock}{\relax}
\providecommand{\bibinfo}[2]{#2}
\providecommand{\BIBentrySTDinterwordspacing}{\spaceskip=0pt\relax}
\providecommand{\BIBentryALTinterwordstretchfactor}{4}
\providecommand{\BIBentryALTinterwordspacing}{\spaceskip=\fontdimen2\font plus
\BIBentryALTinterwordstretchfactor\fontdimen3\font minus
  \fontdimen4\font\relax}
\providecommand{\BIBforeignlanguage}[2]{{%
\expandafter\ifx\csname l@#1\endcsname\relax
\typeout{** WARNING: IEEEtran.bst: No hyphenation pattern has been}%
\typeout{** loaded for the language `#1'. Using the pattern for}%
\typeout{** the default language instead.}%
\else
\language=\csname l@#1\endcsname
\fi
#2}}
\providecommand{\BIBdecl}{\relax}
\BIBdecl

\bibitem{MaGlobecom}
L.~Ma, T.~He, K.~K. Leung, A.~Swami, and D.~Towsley, ``Link identifiability in
  communication networks with two monitors,'' in \emph{IEEE Globecom}, 2013.

\bibitem{GraphTheory2005}
R.~Diestel, \emph{Graph theory}.\hskip 1em plus 0.5em minus 0.4em\relax
  Springer-Verlag Heidelberg, New York, 2005.

\bibitem{Ma13IMC}
L.~Ma, T.~He, K.~K. Leung, A.~Swami, and D.~Towsley, ``Identifiability of link
  metrics based on end-to-end path measurements,'' in \emph{ACM IMC}, 2013.

\end{thebibliography}
\end{document}